# Correction of cell induced optical aberrations in a fluorescence fluctuation microscope

Charles-Edouard Leroux,[1] Alexei Grichine,[2] Irène Wang,[1] Antoine Delon[1]

[1]*Univ. Grenoble 1 / CNRS, LIPhy UMR 5588, Grenoble, 38402, France*
[2]*Univ. Grenoble 1 / INSERM, Institut Albert Bonniot, CRI U823, Grenoble, 38706, France*

*Corresponding author: antoine.delon@ujf-grenoble.fr*

We describe the effect of optical aberrations on fluorescence fluctuations microscopy (FFM), when focusing through a single living cell. FFM measurements are performed in an aqueous fluorescent solution, and prove to be a highly sensitive tool to assess the optical aberrations introduced by the cell. We demonstrate an adaptive optics (AO) system to remove the aberration-related bias in the FFM measurements. Our data show that AO is not only useful when imaging deep in tissues, but also when performing FFM measurements through a single cellular layer.

Fluorescence fluctuation microscopy (FFM) is an ensemble of techniques that in principle allows the absolute measurement of fluorescent molecule concentrations, mobility coefficients and rates of biochemical interactions [1,2,3]. FFM is based on the detection of the fluctuations of the fluorescence signal, and gains in accuracy for a small number of fluorescent molecules inside the observation volume, which is related to the confocal point spread function (PSF). To achieve this condition in biological environments, samples with low concentrations of fluorochromes (nM to µM) are observed with high numerical aperture objectives in a confocal geometry. However, the absolute character of these measurements relies on the assumption that the observation volume is well quantified [4,5]. For most biological applications of FFM, this volume, if not perfectly known, should at least remain stable for comparable measurements at different times or locations. Optical aberrations can prevent a meaningful analysis of FFM measurements by causing sample-dependent distortion of the PSF, which has an impact on both the number of molecules and the characteristic time measured by FFM. We previously demonstrated the utility of an adaptive optics (AO) system for fluorescence correlation spectroscopy (FCS) to greatly reduce this effect in aqueous solutions with various refractive indexes [6], where spherical aberrations caused by index mismatch have dramatic consequences on the measured parameters.

Although most AO developments in microscopy are dedicated to thick specimens such as tissues, where the large amount of aberrations visibly degrades image quality [7], a single cell can already induce significant aberrations, because of the curvatures of the series of interfaces between media of different indices (nucleus, nucleoli, cytosol, etc.). FFM being much more sensitive to aberrations than imaging applications, we show that aberrations caused by a single adhering cell can be very large when observing less than 10 µm above the cell and, consequently, that implementation of an AO system is essential for multicellular layers applications of FFM.

In this Letter, we investigate how passing through a single cell affects FFM measurements, and demonstrate an improved version of our AO system that now uses FFM measurement as optimization metric for automatic aberration correction, without which this work would not has been possible.

We focus our attention on the number of fluorescent molecules N inside the observation volume (one of the outputs of FFM). Assuming that N follows Poisson statistics, it can be estimated as the inverse of the relative fluctuation of the fluorescence signal, when shot noise is negligible. We obtain real-time measurements of N using the mean µ and the variance $\sigma^2$ (evaluated over a few seconds) of the photon count (acquired during a few microsecond binning time): $N = \mu^2/(\sigma^2 - \mu)$. The subtraction of µ in the denominator removes the contribution of shot noise in the overall variance $\sigma^2$ of the signal.

The molecular brightness (defined by the count rate per fluorescent molecule $\mu/N$) is a commonly used metric to quantify the signal-to-noise ratio of a FFM experiment [8], and scales as the Strehl ratio in the presence of optical aberrations of low amplitude [6]. Therefore, we use the measured molecular brightness as a quality metric for modal optimization in our aberration correction scheme, which is similar to the one described for AO based image-sharpening [9]. An important advantage of our method is that it does not require acquiring an image of the sample. Instead, we perform aberration correction in a single point. In this way, the correction is not affected by possible spatial variation of the aberrations. Moreover, since FFM measurements are usually performed in regions of low contrast (where molecules are freely diffusing), image sharpness is only weakly related to optical aberrations [8]. Molecular brightness, in contrast, is a sensitive metric for optical aberration correction in these dilute samples, regardless of the spatial structure.

We have constructed a confocal microscope, which is designed for FFM and equipped with AO. The optical layout is shown in Fig. 1(a). We use a high-speed 97-actuator deformable mirror (DM) (AlpAO, France) for aberration correction. First, the DM is calibrated in a closed-loop scheme using a 32×32 Shack-Hartmann wavefront sensor (SHWFS) (AlpAO), to individually generate 8 Zernike modes (pairs of astigmatisms, comas, trefoils, and spherical aberrations) with root mean square (RMS) amplitudes of ±0.05 µm, the bias values that we use to perform the modal optimization described by Booth et al. [9]. A cycle of aberration correction consists of 3 measurements (of duration 4 s each) per Zernike mode, and is typically repeated 2 to 3 times depending on the level of noise and the amount of initial aberrations. Prior to cell measurements, the microscope inner aberrations are corrected in an aqueous fluorescent solution, by performing two optimization cycles. We then measure the number of molecules ($N_0 \approx 3.4$) and the photon count rate ($\mu_0 \approx 120$ kHz) in the fluorescent solution. The corresponding Zernike modes are defined as default commands to the DM, and typically correspond to a 0.030 µm overall RMS amplitude. The experiment that we describe in this Letter is illustrated in Fig. 1(b). We perform FFM measurements through mouse embryo fibroblast cells, in the same fluorescent solution that we use to correct the microscope aberrations. Doing so allows us to investigate the optical effect induced by the cell, as compared to the nominal measurements in the cell-free fluorescent solution. The FFM measurements µ and N are normalized with the cell-free measurements: $\tilde{\mu} = \mu/\mu_0$ and $\tilde{N} = N/N_0$. To minimize the amplitude of the initial aberrations when starting cell experiments, we use neighboring wells in the same chambered coverslip (Nunc Labtek) for system aberrations correction and cell measurements. In the following, we show typical FFM data obtained with two cells that spread differently on the glass substrate: a spherical cell1 (Fig. 2(a)) and a flat cell2 (Fig. 2(b)).

We show in Figs. 2(c-f) FFM measurements as a function of the focus position z, which we define relative to the apex of the cell (z = 0 µm). We first acquire measurements with the DM set to its default commands (dotted curves). At the cell-water interface, the number of molecules is close to the cell-free values ($\tilde{N} \approx 1.1$ at z = 0 µm for the two cells, Figs. 2(c) and 2(d)). This observation suggests that the PSF is not significantly distorted by optical aberrations. However, there is a loss of photons ($\tilde{\mu} \approx 0.7$ at z = 0 µm, dotted curves of Figs. 2(e) and 2(f)), which is presumably not related to optical aberrations for two reasons: i) there is no significant increase of µ after aberration correction (data not shown) and ii) µ is generally less sensitive to optical aberrations than N. This observation holds true for any FFM measurement in a dilute sample, and is related to the fact that the image of a diffuse object is only weakly affected by optical aberrations [8]. In the limit of a very large detector, µ is independent of aberrations because it scales as the incoming optical power. This loss of photons measured at z = 0 µm is presumably due to light scattering in the cell. To observe the effect of optical aberrations, it is necessary to focus a few micrometers above the cell. The number of molecules increases to a maximum (cell1: $\tilde{N} \approx 7.5$ at $z_{max}$ = 8 µm, cell2: $\tilde{N} \approx 3.2$ at $z_{max}$ = 13 µm). We interpret $z_{max}$ as the focus position for which the entire wavefront covers the cell, and is therefore the most aberrated. Focusing above this position reduces the optical aberrations, because the marginal rays do not propagate through the cell. At z = $z_{max}$, the photon count is minimum (cell1: $\tilde{\mu} \approx 0.5$, cell2: $\tilde{\mu} \approx 0.6$), but the relative decrease compared to its value at the cell-water interface is much smaller than for $\tilde{N}$.

Aberration correction at z = 8 µm (cell1) and z = 13 µm (cell2) reduces the number of molecules to $\tilde{N} \approx 1.5$ (factors 5 and 2 for cell1 and cell2 respectively, see the solid curves of Figs. 2(c) and 2(d)). Larger aberrations are corrected for cell1 (overall aberration is 0.11 µm RMS, including 0.09 µm of spherical aberration $Z_{4,0}$) than for cell2 (0.07 µm RMS, including $Z_{4,0}$ = 0.05 µm). It is clear from the curves of Figs. 2(c-f) that the optimal wavefront strongly depends on the focus position. The aberration correction performed at 8 µm and 13 µm above cell1 and cell2 strongly distorts the PSF at other focus positions. In particular, it increases $\tilde{N}$ at z = 0 µm: $\tilde{N} \approx 7.5$ and $\tilde{N} \approx 4$, for cell1 and cell2 respectively (solid curve of Figs. 2(c) and 2(d)).

It is also possible to see the effects of optical aberrations on the number of molecules using the raster image correlation spectroscopy (RICS) technique [10,11]. This technique consists in analyzing the spatio-temporal correlation of confocal images acquired at a relatively slow scan speed (5nm/µs) and with a pixel size much smaller than the PSF, so that the probability of detecting the same diffusing molecule at two consecutive pixels is non-zero. Each sub-region of the original confocal images, typically 32x32 to 128x128 pixels, provides a pixel in the N(x,y) 2D map. A RICS analysis in a fluorescent solution provides homogeneous maps of N(x,y) in the absence of instrumental artifacts such as optical aberrations.

As in non-scanning FFM techniques (such as FCS), N(x,y) is proportional to the observation volume. Fig. 3 shows the effect of optical aberrations on N(x,y) maps recorded above cell1 (Fig. 3 (a)) and cell2 (Fig. 3 (b)), at z = $z_{max}$ (with the DM set to the default commands). The "shadow" of each cell is visible in these maps because of optical aberrations, but barely appears on the raw confocal images (data not shown). At the periphery of the maps, $\tilde{N}$ is minimal ($\tilde{N} \approx 1.2$ for both cells), because a smaller fraction of the beam propagates through the cell. $\tilde{N}$ is larger at the centre of the cells (cell1: $\tilde{N} \approx 7$, cell2: $\tilde{N} \approx 3$, as in Figs. 2(c) and 2(d) without AO). After aberration correction, the RICS analysis shows the reduction of $\tilde{N}$ in the central region of the two maps (Figs. 3(c) and 3(d)). For cell1, the region for which aberration correction is beneficial is very small, and corresponds to the center of the cell. This region is larger for cell2, and is elongated along the direction of the cell. This latter observation illustrates the idea that the full benefit of AO in a microscope is only achieved when the aberration correction can be updated across the field of view, as illustrated in Ref [12].

To conclude, we demonstrated that the optical aberrations introduced by a single living cell can have a significant impact on FFM, and that AO is a promising technique for performing robust FFM measurements in complex biological samples such as multicellular layers. FFM provides a metric that is directly related to optical aberrations, even in dilute regions of the sample. This feature makes FFM potentially useful for aberration correction in other microscopy techniques. The measurement of molecular brightness is local, and could therefore be used as optimization metric for aberration correction at different points of the field of view. Doing so would require some work to understand how optical aberrations spatially change in the sample.

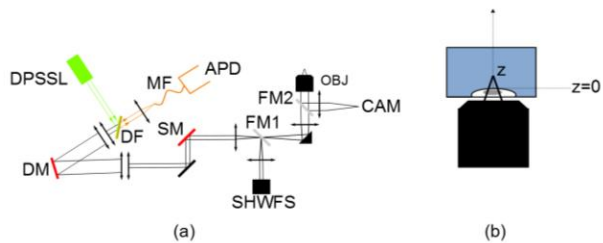

Figure 1: (a) The optical layout. DPSSL: 561 nm diode-pumped solid state laser (Cobolt); DM: 97 actuator deformable mirror (AlpAO); OBJ: 63×/1.2 water immersed microscope objective (Zeiss); SM: 3 mm X/Y galvanometric mirrors (Cambridge Technology); APD: single photon counting avalanche photodiode (PerkinElmer); MF: 1xAiry multimode fiber; DF: 600 nm long-pass dichroic filter (Chroma); SHWFS : 32×32 Shack-Hartmann wavefront sensor (AlpAO). CAM : wide field camera (ANDOR sCMOS Zyla); FM1: flip mirror for DM calibration; FM2: flip mirror for transmission microscopy. (b) Schematic of the experiment. The z = 0 focus position corresponds to the apex of the cell. FFM measurements are carried out in a fluorescent solution (sulforhodamineB at a 15 nM concentration).

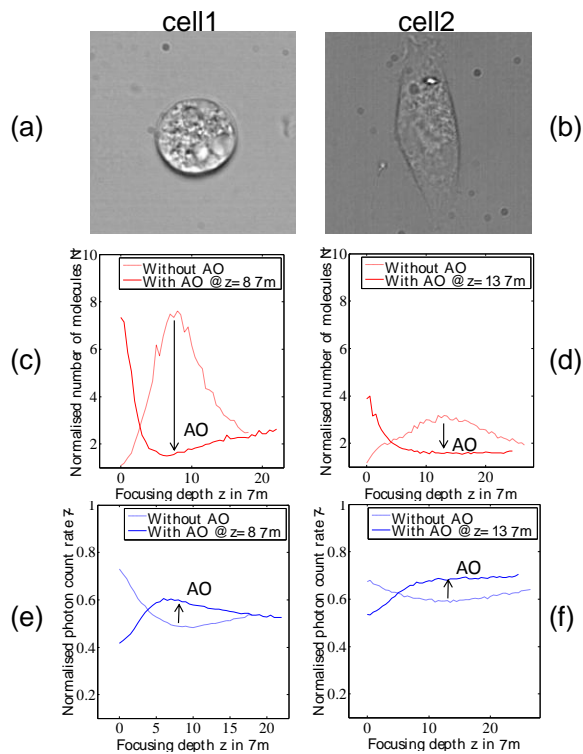

Figure 2. (a/b) Wide field images of cell1/cell2 over a 50×50 µm field of view, observed with transmission microscopy. FFM measurements are carried out at the centre of the field of view. (c/d) Number of molecules function of the focusing distance above the center of cell1/cell2, before (dotted line) and after (solid line) aberration correction. (e/f) Photon count rate above the center of cell1/cell2, before (dotted line) and after (solid line) aberration correction.

The authors thank Catherine Souchier and Olivier Destaing for helpful discussions and cell preparation. This work is supported by the ANR-11-EMMA-0036 research program.

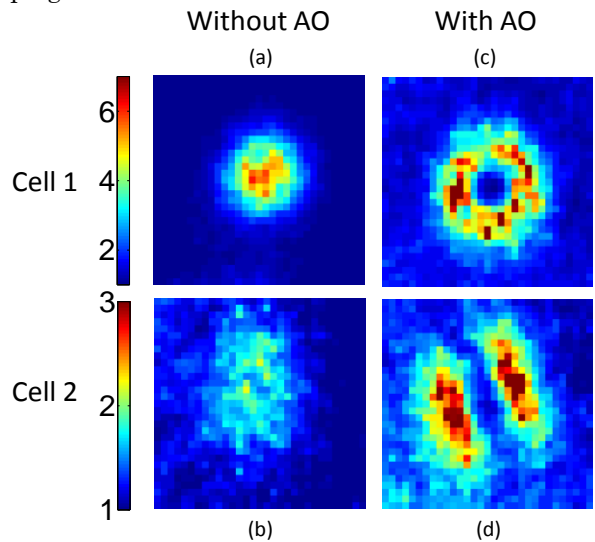

Figure 3. N(x,y) maps of the number of molecules over a 50×50 µm field of view, using the RICS technique. (a/c) Cell1 without/with AO. (b/d) Cell2 without/with AO.